\documentclass{article}
\usepackage{amsmath,mathrsfs,bm}
\usepackage{natbib}   %[authoryear]
\usepackage{graphicx}
\usepackage{hyperref}

\title{Reading tea leaves? Polygenic scores and differences in traits among groups.}

\author{Graham Coop}
\begin{document}

\maketitle

\begin{abstract}
In the past decade, Genome-Wide Association Studies (GWAS) have delivered an increasingly broad view of the genetic basis of human phenotypic variation. One of the major developments from GWAS is polygenic scores, a genetic predictor of an individual's genetic predisposition towards a trait constructed from GWAS.  The success of GWAS and polygenic scores seems to suggest that we will soon be able to settle debates about whether phenotypic differences among groups are driven in part by genetics. However, answering these questions is more complicated than it seems at first glance and touches on many old issues about the interpretation of human genetic variation. In this perspective piece I outline the ways in which issues of causality, stratification, gene-by-environment interactions, and divergence among groups all complicate the interpretation of among-population polygenic score differences.
\end{abstract}

%\mde{delete "and contentious," kind of redundant w/ "debate"} 
Debates about the contribution of genetics to differences among groups have a long and contentious history. We have known for a long time that nearly all traits are partially heritable, meaning that variation in the traits is due both to genetic differences as well as environmental ones. The heritability of traits within groups has led some observers---both scientists and non-scientists---to conclude that between-group differences in average traits can also be explained in part by genetics, approximately in proportion to their heritability. Yet if a trait is highly heritable within a population, it does not follow that differences between groups are due to genetics—instead environmental and cultural differences could be the primary driver of between-population differences \citep{feldman1975heritability}.  %% could cite https://science.sciencemag.org/content/190/4220/1163
 
Recently, the field of genetics has made progress in identifying regions of the genome (mostly single nucleotide polymorphisms, SNPs) that are associated with differences among individuals within a population, using genome-wide association studies (GWAS). GWAS have found SNPs associated with a dizzying array of traits. Concurrently, sophisticated methods have emerged to estimate the combined phenotypic effect of many genetic loci citep{vinkhuyzen2013estimation}. The success of GWAS seems to suggest that we will soon be able to settle debates about whether average phenotypic differences among populations are driven in part by genetic differences. Answering this question is harder than it seems at first glance, however. In this perspective, I will explain some of the complications, including how stratification, gene-by-environment interactions, and correlations among SNPs can make it difficult or in some cases effectively impossible to use GWAS to understand differences among populations.

Some of these complications are perhaps best illustrated with a toy example. Say we perform a GWAS of the amount of tea that individuals in the UK drink \citep[e.g. in the UK Biobank, ][]{taylor2018associations,cole2019comprehensive}. On the basis of this tea-drinking GWAS, a person might argue that we could learn about France-UK differences in tea consumption by counting up the average number of alleles for tea preference that individuals in the UK and France carry. If the British, overall, are more likely to have alleles that are associated with increased tea consumption than the French, then he might claim that we have demonstrated that the difference between French and UK peoples’ preference for tea is in part genetic. Our protagonist would assure us that these alleles are polymorphic in both countries  and that both environment and culture plays an important role. He would further reassure us that there will be an overlapping distribution of numbers of tea-drinking alleles in both countries, so he's not saying that all British people drink more tea than all French people for genetic reasons. He'll tell us he's simply interested in showing that the average difference in tea consumption is partly due to genetic differences between groups.
 
At face value, this argument seems scientifically plausible; if there are alleles for tea preference, then to determine whether the British peoples’ love of a good cuppa tea is genetic, we could simply count these alleles up and compare them to the average allele counts in France. Adding up these tea preference alleles for individuals is one way of calculating an individual's "polygenic score". Polygenic scores can be thought of as an attempt at the prediction of people’s trait values computed from genotype data. There are several ways of calculating polygenic scores, and they have a range of potential uses. For example, people have done GWAS for risk of coronary artery disease  and constructed polygenic scores in the hope of enabling early detection and preventive care \citep{khera2018genome}. Currently, these polygenic scores tend not to explain much of the variation in a given trait, but the size of studies is increasing, and predictions based on polygenic scores will likely become more accurate within populations \citep[but see][]{mostafavi2019variable}.

Polygenic scores constructed using GWAS information from a single population are expected to differ among populations. The allele frequency at every locus will vary among populations because of genetic drift--- the compounding of chance variation in which alleles are transmitted from generation to generation---leads allele frequencies among populations to diverge over time. If natural selection acts on the loci differently in the two populations, it will also cause allele frequencies to differ, but genetic drift is sufficient to cause allele frequency differences and will occur regardless of whether there is selection on the trait. Since a polygenic score is just a weighted sum of allele counts, it will also vary among populations in ways that are reasonably well understood in the case of genetic drift \citep{Berg:2014bs}. 

Importantly, the existence of genetic differences among populations does not imply that observed differences in phenotype among populations are the result of genetics. It could be that French people tend to have higher polygenic scores for tea-consumption than the British, but that this genetic predisposition is hidden or counteracted by cultural influences. For example, perhaps British people on average find bitter (tannin) tastes slightly less palatable than French people, but this influence is overridden by the culture of tea drinking in the UK. It is a simple point that genetics and environment can act in opposite influences on a phenotypic difference between populations, yet people often assume that differences in PGS should somewhat line up with observed phenotypic differences as if that were not the case.
 
Beyond the fact that environment and culture can counteract or overwhelm the influence of genetics, there’s another, more subtle problem: polygenic scores are not strong statements about differences in the contribution of genetics to phenotypic variation among groups. The issue is that GWAS studies do not point to specific alleles \emph{for} tea preferences, only to alleles that are associated with tea preference in the  set of environments experienced by people enrolled as participants in the GWAS in the UK Biobank. Similarly, as geneticists, we often talk about height alleles and loci obtained from GWAS. But these are not alleles \emph{for} height, but simply alleles that are associated with differences in height within a population that was studied in a GWAS. There's no guarantee that alleles mapped within populations will affect the trait in the same way in other populations and environments, nor (even if they do) that they will explain differences between populations. 

Complex traits are just that---complex.  Most traits are incredibly polygenic, likely involving tens of thousands of loci. These loci will act via many pathways \citep{boyle2017expanded}, mediated by interactions with many environmental and cultural factors.  Some of our tea-GWAS SNPs may well be enriched near olfactory receptors and genes expressed in relevant parts of brain, and some may overlap with SNPs associated with caffeine sensitivity \citep{taylor2018associations}. But the majority may not---many may fall near genes with no simple connection to tea drinking. For example, how thirsty you get may play into their beverage choice, thus genetic variants in many physiological pathways may be weakly associated with tea drinking.  The rare cases where we can confidently make a specific causal connection to a gene and through a causal pathway all the way to phenotype may explain so little of the variance that, while they may provide important clues to biology, they often won’t allow us to state a general causal mechanism that underlies our genetic predictions. This should not be taken as an anti-GWAS position. We have learned a lot of new biology from GWAS \citep{visscher201710}, and doubtless will learn a lot more over the coming years. But GWAS are not a complete solution to understanding the causes of variation, especially variation among populations and across environments \citep{rosenberg2018interpreting}. Let’s see some reasons why.

\subsection*{Population stratification} 
One long-standing concern in GWAS is the potential confounding of environmental and genetic variation due to population stratification.  To co-opt a classic illustrative example of the problem \citep{lander1994genetic},consider performing a GWAS of the phenotype "drinks strong tea" in Paris. Any allele at slightly higher frequency in English immigrants in Paris than in other Parisians could be spuriously correlated with tea drinking. These stratification issues will arise whether the cause of British preference for tea drinking is environmental or partially genetic \citep{vilhjalmsson2012nature}. Such stratification issues may replicate across seemingly independent datasets. There is no such thing as a truly independent sample of humans--just samples with varying degrees of relatedness thus stratification issues may replicate even across seemingly quite different samples.

Concerns about stratification drove human GWAS to be implemented primarily in what were characterized as relatively genetically homogeneous populations, with the hope that genotypes would be reasonably randomized across environmental and genetic backgrounds within that population. To mitigate against any remaining confounding by stratification, a broad range of statistical techniques were developed  \cite{price2010new,bulik2015ld}. These advances have likely greatly reduced some of the issues of stratification in modern GWAS, notably for the strongest associations. However, even a small amount of residual stratification can potentially cause serious problems for the interpretation of polygenic scores, as this subtle bias is compounded across the loci used to construct a polygenic score. This problem does not boil down to false-positive genetic associations, because even true positives can have effect size estimates that are subtly biased by stratification. 

One vivid example is offered by recent issues with height polygenic scores. Starting in 2012 \citep{Turchin:2012dy}, a number of studies identified a seemingly strong difference in polygenic scores for height between Northern and Southern Europeans \cite{Berg:2014bs,berg2017polygenic,Mathieson:2015cy,Robinson:2015ki,Zoledziewska:2015do,Racimo18,guo2018}. To this end, they relied on scores that were constructed from effect sizes estimated by the GIANT GWAS, a meta-analysis of various European ancestries samples, and seemingly replicated their results in various ways. However, when a number of studies tried to replicate the European height results using GWAS effect sizes from the UK Biobank \cite{bycroft2018uk}---a large, independent sample---they found that the previously observed patterns did not replicate \citep{berg2019reduced,sohail2019polygenic} and upon further inspection, found problems with population stratification in the original GIANT height GWAS and replication efforts. Many of the top hits did replicate between the GIANT and UK Biobank studies, so arguably the GIANT study fulfilled its main aim. However, GIANT was not a reliable basis for building PGS, which compounds many subtly biased effect size estimates. Other papers have begun to highlight potentially related issues \cite{kerminen2018geographic,haworth2019apparent}, demonstrating that stratification may be a serious confounding issue for polygenic scores at even smaller geographic scales. 

For many traits, there may be no bright line between population stratification and `real' biological effects \cite{lawson2019population,haworth2019apparent,belsky2019genetics}. People in better health are more likely to move long distances within the UK \citep{brimblecombe2000migration}. If people tend to move to areas with less tea drinking per capita, then alleles associated with overall health wd then become alleles associated with decreased tea drinking. Adjusting for genome-wide relatedness in the GWAS would likely not make this effect go away---the induced correlations are beyond those expected due to stratification due to genome-wide relatedness. Yet this too is a form of stratification, one that may not replicate among groups. 

\subsection*{Gene-by-environment interactions (G x E)}
 The effect of an allele on any given phenotype is always measured in the context of a particular set of environments. This issue is not new: debates over the meaning of heritability and the genetics in the context of environmental variation stretch back to the dawn of quantitative genetics and debates between Hogben and Fisher \citep{tabery2008ra}. These issues are particularly difficult in humans, as we cannot raise humans in laboratory environments or conduct randomized environments. Our behavioural, cultural and societal practices will influence the ways in which genetic variants impact phenotypic variation.
 
For example, there are cultural differences between the UK and France in whether milk is taken with tea, in the types and quality of tea drunk, and in the prominence of coffee. Do polymorphisms associated with bitter taste sensitivity lead to different beverage outcomes in the two countries? What role do parents, siblings, and peers play in shaping one's choice of hot drinks, and how do these interactions differ between countries? The genotypes of other people can indirectly shape one's phenotype by shaping one's environment \citep{kong2018nature}. Presumably, all of these differences, and many others,  could mean that the genetic basis of variation in tea drinking will differ between France and the UK. Therefore, the loci that influence tea drinking in the UK could be somewhat different from those underlying differences in tea drinking in France.
 
Even if we perform tea-drinking GWAS in multiple countries, we may not be able to circumvent the issues of GxE. Suppose after our GWAS for tea drinking in the UK and France, we find found that the genetic basis of the trait within both countries to be correlated. What would be a high enough correlation to constitute evidence of a genetic difference in phenotypic preferences between countries? Observing a high genetic correlations in GWAS results amongst groups does not mean that polygenic predictions are necessarily portable between groups \citep[even within groups with very similar ancestries][]{mostafavi2019variable}.  Moreover, even if the polygenic score explained a lot of the variance within each country, it may not explain much of the difference between the countries. As one example: suppose, hypothetically, that people who care greatly about their weight are more likely to drink tea (e.g., as compared with soft drinks); then alleles that are correlated with body mass index (BMI) in the UK Biobank will be alleles predicted to predispose one to tea drinking. These loci may be reliably associated with BMI and tea drinking in both the UK and France. Yet a difference in the frequency of loci associated with BMI between the UK and France would not imply that differences in tea drinking preferences among countries result from genetics. Suppose for example that an individual's preference for tea is not influenced by their absolute BMI, but rather by their relative BMI within a country. In this scenario, a polygenic score could be predictive of individual phenotypes within multiple countries but have little predictive power in explaining differences among those countries.
 
Without a thorough understanding of the causal biological and cultural mechanisms by which GWAS SNPs interact with the range of environments encountered by individuals, it may be hard to rule out GxE as a serious confounder of inferences of polygenic scores across populations. These problems do not just exist among populations. There's systematic variation across the UK in opinions about how milky tea should be, and whether milk should be added before or after the tea \citep[a matter of intense debate][]{fisher1937design,Orwell:1946,salsburg2001lady}. Tastes in tea and use of sugar vary across socio-economic groups with UK \citep{CuppaTheGrocer}. What phenotype, and which set of environments, are we learning about when we examine polygenic scores? 
 
 %https://www.thegrocer.co.uk/buying-and-supplying/categories/hot-beverages/revealed-what-your-cuppa-says-about-you/524797.article
 
\subsection*{We don’t have the functional genetic markers}
\begin{figure}
\begin{center}
\includegraphics[width= 0.5 \textwidth]{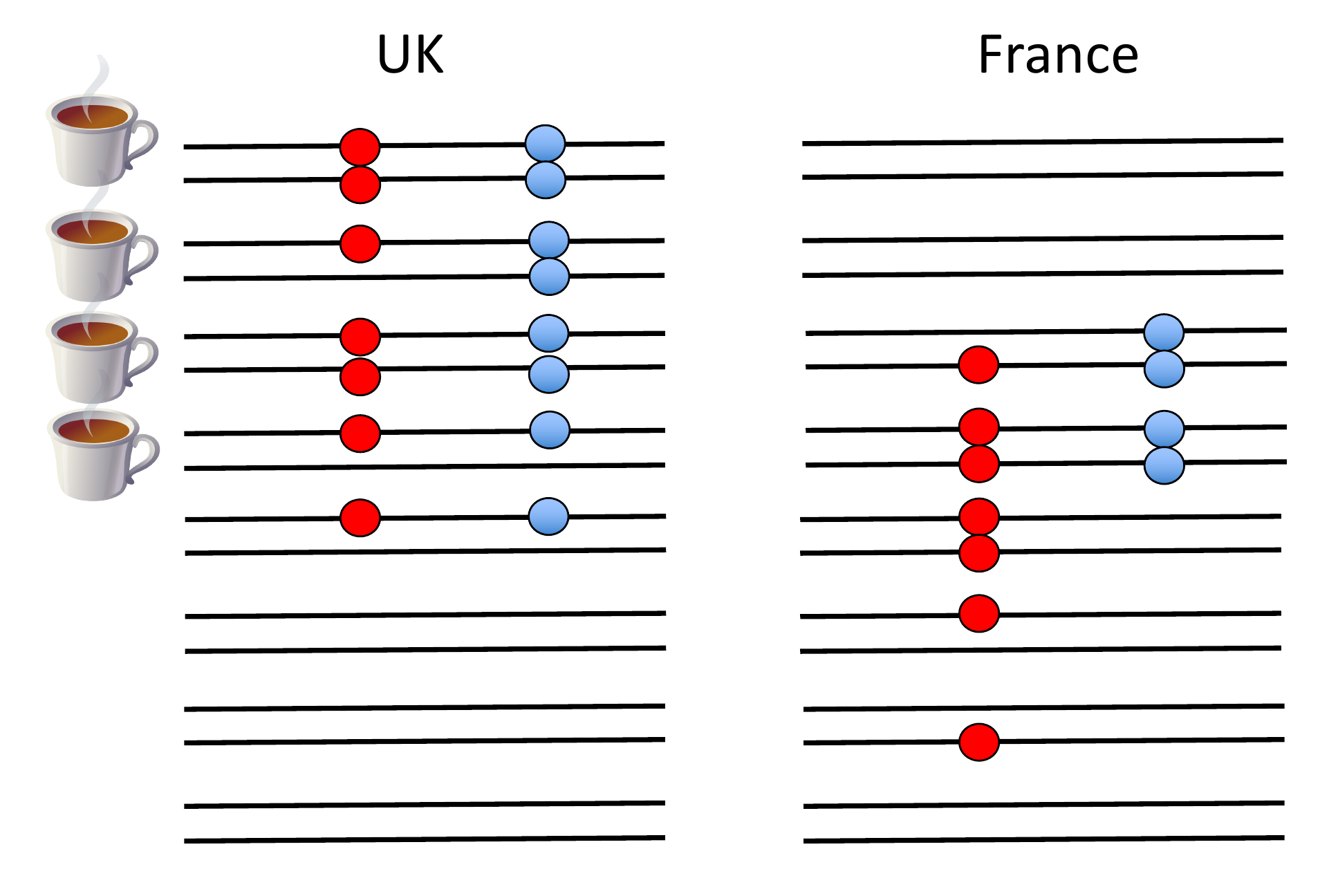}
\end{center}
\caption{Each person has two chromosomes (horizontal black lines) and in this small stretch of the genome there are two loci (red and blue SNPs), the alleles of which are indicated by the presence/absence of a filled circle. Whether an individual drinks a lot of tea is indicated by the tea cup next to the individual (public domain \href{https://publicdomainvectors.org/en/free-clipart/Coffee-or-tea-sticker-vector-drawing/22666.html}{tea cup}).  Obviously, this sample size is too small, but it serves for illustration.} \label{fig:tea_LD}  
\end{figure}

A third major hurdle that we face in understanding polygenic scores is that we do not know the loci that are functionally important for trait variation, only loci that are statistical proxies for them---sometimes called tag SNPs---that will be nearby in the genome. (Technically the SNPs used to construct polygenic scores are in linkage disequilibrium with the functional loci---meaning that genotypes at the tag SNP are correlated with genotypes at the functional locus---but unlikely to be the functional loci themselves.) To understand this point, look at the example below (Figure \ref{fig:tea_LD}). On the left is a cartoon of people from the UK. Both of the filled circle alleles appear to be associated with tea drinking. ( However, only one of them is the functional SNP predisposing people toward tea drinking; the other SNP just happens to be associated because the mutation there arose at a similar point in history on the same genetic background. If we guess that the blue allele is the functional one, we would predict that French people have a slightly weaker preference to tea on the basis of this allele. But if we guess the red allele is the functional one, we would predict that the UK and France have very similar tea drinking habits on the basis of this locus.
 
What's happened here is that the correlation between the alleles at the two loci have changed due to different histories of recombination and genetic drift. Now, such a strong change in the correlation of loci is unlikely between two countries, such as Britain and France, that share so much of their genetic history. However, it is a serious problem when comparing populations that have been more distant from each other for a longer period of time. The fact that the correlation between any two SNPs changes over evolutionary time is a major candidate explanation for why polygenic scores lose predictive ability as we move to populations that have been isolated for more of their history from the population in which the GWAS was conducted \citep{carlson2013generalization,marigorta2013high,martin2019clinical}. For highly polygenic traits, the associations we find may partially reflect collections of loosely linked SNPs, whose relationship may be subject to change even between closely related populations. 
 
  %https://publicdomainvectors.org/en/free-clipart/Coffee-or-tea-sticker-vector-drawing/22666.html
 
A second, more subtle force can decrease the predictive validity of polygenic scores. Assortative mating among individuals can drive rapid changes in the SNPs associated with a trait. For example, if people who drink more tea tend to have children with taller people, this pattern of assortative mating can cause greater height and tea drinking to become associated \cite[i.e., assortative mating can generate genetic correlations][]{gianola1982assortative}. In other words, height-increasing alleles will be associated with tea drinking because the offspring of tea-drinking/tall couples will have alleles associated with both tea drinking and height. Even after assortative mating has stopped, these effects can persist for a few generations as they are slowly broken down by Mendelian segregation and recombination, making them potentially hard to rule out.  Such associations need not hold in other populations, however, if they have not experienced similar patterns of assortative mating. Therefore, sets of loci that contribute to trait variation via genetic correlations may change rapidly across environments or populations due to changes in patterns of assortment.
 
\subsection*{We will not map all of the alleles influencing trait differences among populations}
Other things equal, the statistical power of GWAS is higher for alleles that are at intermediate frequency in the GWAS population than for alleles that are at low or high frequency. The functional variants contributing to a trait will differ in frequency among populations due to genetic drift and selection. Therefore, GWAS conducted in one population will miss many of the loci contributing to phenotypic variation in other populations. This may not be much of a problem for comparing the UK and French population, as allele frequencies are very similar in the two countries. However, it is potentially a much bigger problem in comparing more distant populations.
 
A case in point is the genetics of skin pigmentation. The variants that were mapped within European populations, though important in Europe, explain little of the variation in skin pigmentation worldwide. Even variants that explain lighter European skin pigmentation do not explain lighter skin pigmentation in East Asians \citep[e.g. see ][]{Norton:2006kpa,Edwards:2010kda,adhikari2019gwas}.  Loci important for explaining skin-pigmentation variation worldwide were missed by studies focused on non-African populations and only found as variation within Africa began to be explored \citep{crawford2017loci,martin2017unexpectedly}. Furthermore, our understanding of the evolutionary history of skin pigmentation in Europe is undergoing major revision in the light of ancient DNA \citep{olalde2014derived}.  This history of major shifts in our understanding of the genetics and evolutionary history of skin pigmentation suggests that bold claims about other traits, based on incomplete evidence, may not stand the test of time.
 
In the coming decade, we will likely uncover a surprising amount of heterogeneity in the alleles controlling trait variation worldwide. Based on genetic drift alone, we should expect as much: the alleles that explain most variance in populations of European ancestry  and those which explain the most variance in those of East Asian ancestry will not be the same, because allele frequencies drift over time. Also as a result of allele frequency change at many loci, across populations, epistatic relationships among loci may also change in unpredictable ways, confounding cross-population predictions. These problems of different alleles contributing to  traits in different populations will be compounded for traits subject to natural selection (in addition to genetic drift). Whether traits are subject to stabilizing selection or directional selection (shared or divergent), selection will drive more rapid turnover in the loci contributing to trait variation among populations \citep{durvasula2019negative}.
 
Again, one can hope to address these issues by performing GWAS in multiple worldwide populations, but the reality is that we will have a European-biased view of genetic variation for some time to come, simply because of the size of the studies in these populations dwarfs those that can currently be done elsewhere.

\subsection*{Conclusion}

Undoubtedly the coming decades of human genomics will see advances in the identification of functional loci, the size of GWAS performed worldwide, and in the statistical methodologies used to understand trait variation. 
The conceptual issues and pitfalls discussed here are fascinating problems in their own right. 
Working to understand the interplay of GxE along with direct and indirect effects of genetics promises to be a rich seam of work for many years to come. Interpreting differences in mean polygenic scores poses many challenges, but some promises. Understanding cross-population differences in genetics may allow us to better understand the contribution of genetics to differences in disease prevalence among populations \citep[although many differences likely reflect healthcare and environmental disparities][]{williams2005social,williams2009discrimination}. While the ability to use GWAS to learn about whether natural selection has driven differences in the genetic architecture of human traits among populations promises to open up new intersections between anthropology, genetics, and archaeology; for example in understanding the role of environmental change and genetics in shaping changes in height and body proportions during the Neolithic \citep{cox2019genetic}. There is also no doubt that we will come to understand much more about human variation. However, our ability to perform GWAS to identify loci underlying variation in traits among individuals vastly outstrips our ability to understand the causal mechanisms underlying these differences. In many cases, genetic contributions may not be separable from environmental and cultural differences. Certainly making a case for the relative importance of genetics in explaining differences among populations will involve a lot more work than simply counting up the number of tea preference alleles in populations and seeing how the averages differ.
 
Despite these complications, we are poised to see many partial, incomplete (and in some cases initially downright incorrect) stories about the genetics of population differences in traits. Already, we thought we knew something about the evolution of polygenic height scores among European populations, only to find that much of this result was driven by stratification-- and that’s for height, an easily measured and well-studied trait. Applied to other and more fraught traits, this patchy understanding of the contribution of genetics to phenotypic differences will be fertile ground for misleading claims.

At the core of this issue is an even more fundamental disconnect between talk of polygenic scores and what some people seem to think they might learn from this kind of research. Even if we could attribute some proportion of the phenotypic difference among populations to differences in polygenic score, such a result won’t address the question that many have in mind in asking whether a difference is ``genetic.” Saying a phenotypic difference among individuals is genetic is often taken to imply that it is immutable or unavoidable. Yet a difference in genetic predispositions are neither immutable or “natural”, as many people have highlighted \citep{penrose1955heredity}. Many phenotypes where genetics plays a role are modifiable, and presumably many "genetic" trait differences among people are environmentally mediated. Without at least some working knowledge of causal mechanisms underlying the action of the genetic variation contributing to a trait, we may often not know how environment and culture shape the actions of these variants, nor how changes in these factors may modify any role played by genetics. Even if our tea polygenic scores were strongly predictive within and among populations, cultural changes, e.g., a Europe-wide health food craze for drinking tea with dinner, could presumably stand these results on their head. Will taking tea with a meal moderate the role of caffeine-sensitivity SNPs; will exercise-conscious people now drink more tea? Will we know enough about the interaction of culture and genetics to predict this? If we do not, the statement that a difference in polygenic scores plays a role in explaining a difference in phenotypes among populations will shed little light on how we as individuals or societies should view that difference. But will these critical subtleties be lost in the public’s understanding of results based on polygenic scores? Will such results be wrongly taken as supporting genetic determinism about human variation? 
%\mde{Perhaps ending on some stronger statement?}

% If your first paragraph (i.e. with the \dropcap) contains a list environment (quote, quotation, theorem, definition, enumerate, itemize...), the line after the list may have some extra indentation. If this is the case, add \parshape=0 to the end of the list environment.
{\bf Acknowledgements.} I thank Jeremy Berg, Vince Buffalo, Nancy Chen, Doc Edge, Emily Josephs, Molly Przeworski, and Jenny Tung for helpful discussions and/or feedback on earlier drafts. This work was supported in part by the National Institute of General Medical Sciences of the National Institutes of Health (NIH R01 GM108779). % Display the acknowledgments section

%\subsection*{References}

%\begin{figure}%[tbhp]
%\centering
%\includegraphics[width=.8\linewidth]{frog}
%\caption{Placeholder image of a frog with a long example caption to show justification setting.}
%\label{fig:frog}
%\end{figure}

%\begin{SCfigure*}[\sidecaptionrelwidth][t]
%\centering
%\includegraphics[width=11.4cm,height=11.4cm]{frog}
%\caption{This caption would be placed at the side of the figure, rather than below it.}\label{fig:side}
%\end{SCfigure*}

% Bibliography
\bibliographystyle{genetics}
\bibliography{Grahams_bib,library,library2,library3}

\begin{thebibliography}{}

\bibitem[\protect\citeauthoryear{Adhikari, Mendoza-Revilla, Sohail,
  Fuentes-Guajardo, Lampert, Chac{\'o}n-Duque, Hurtado, Villegas, Granja,
  Acu{\~n}a-Alonzo, {\it et~al.}}{\textsc{Adhikari} {\em
  et~al.\@}}{2019}]{adhikari2019gwas}
\textsc{Adhikari, K.}, \textsc{J.~Mendoza-Revilla}, \textsc{A.~Sohail},
  \textsc{M.~Fuentes-Guajardo}, \textsc{J.~Lampert}, \textsc{J.~C.
  Chac{\'o}n-Duque}, \textsc{M.~Hurtado}, \textsc{V.~Villegas},
  \textsc{V.~Granja}, \textsc{V.~Acu{\~n}a-Alonzo}, and \textsc{others}, 2019\
  \ A GWAS in Latin Americans highlights the convergent evolution of lighter
  skin pigmentation in Eurasia.
\newblock Nature communications~{\em 10\/}(1)\textbf{:}\ 358.

\bibitem[\protect\citeauthoryear{Belsky, Caspi, Arseneault, Corcoran, Domingue,
  Harris, Houts, Mill, Moffitt, Prinz, {\it et~al.}}{\textsc{Belsky} {\em
  et~al.\@}}{2019}]{belsky2019genetics}
\textsc{Belsky, D.~W.}, \textsc{A.~Caspi}, \textsc{L.~Arseneault},
  \textsc{D.~L. Corcoran}, \textsc{B.~W. Domingue}, \textsc{K.~M. Harris},
  \textsc{R.~M. Houts}, \textsc{J.~S. Mill}, \textsc{T.~E. Moffitt},
  \textsc{J.~Prinz}, and \textsc{others}, 2019\ \ Genetics and the geography of
  health, behaviour and attainment.
\newblock Nature human behaviour\textbf{:}\ 1.

\bibitem[\protect\citeauthoryear{Berg and Coop}{\textsc{Berg} and
  \textsc{Coop}}{2014}]{Berg:2014bs}
\textsc{Berg, J.~J.} and \textsc{G.~Coop}, 2014, August){A population genetic
  signal of polygenic adaptation.}
\newblock PLOS Genetics~{\em 10\/}(8)\textbf{:}\ e1004412.

\bibitem[\protect\citeauthoryear{Berg, Harpak, Sinnott-Armstrong, Joergensen,
  Mostafavi, Field, Boyle, Zhang, Racimo, Pritchard, {\it
  et~al.}}{\textsc{Berg} {\em et~al.\@}}{2019}]{berg2019reduced}
\textsc{Berg, J.~J.}, \textsc{A.~Harpak}, \textsc{N.~Sinnott-Armstrong},
  \textsc{A.~M. Joergensen}, \textsc{H.~Mostafavi}, \textsc{Y.~Field},
  \textsc{E.~A. Boyle}, \textsc{X.~Zhang}, \textsc{F.~Racimo}, \textsc{J.~K.
  Pritchard}, and \textsc{others}, 2019\ \ Reduced signal for polygenic
  adaptation of height in UK Biobank.
\newblock eLife~{\bf 8}\textbf{:}\ e39725.

\bibitem[\protect\citeauthoryear{Berg, Zhang, and Coop}{\textsc{Berg} {\em
  et~al.\@}}{2017}]{berg2017polygenic}
\textsc{Berg, J.~J.}, \textsc{X.~Zhang}, and \textsc{G.~Coop}, 2017\ \
  Polygenic adaptation has impacted multiple anthropometric traits.
\newblock BioRxiv\textbf{:}\ 167551.

\bibitem[\protect\citeauthoryear{Boyle, Li, and Pritchard}{\textsc{Boyle} {\em
  et~al.\@}}{2017}]{boyle2017expanded}
\textsc{Boyle, E.~A.}, \textsc{Y.~I. Li}, and \textsc{J.~K. Pritchard}, 2017\ \
  An Expanded View of Complex Traits: From Polygenic to Omnigenic.
\newblock Cell~{\em 169\/}(7)\textbf{:}\ 1177--1186.

\bibitem[\protect\citeauthoryear{Brimblecombe, Dorling, and
  Shaw}{\textsc{Brimblecombe} {\em et~al.\@}}{2000}]{brimblecombe2000migration}
\textsc{Brimblecombe, N.}, \textsc{D.~Dorling}, and \textsc{M.~Shaw}, 2000\ \
  Migration and geographical inequalities in health in Britain.
\newblock Social science \& medicine~{\em 50\/}(6)\textbf{:}\ 861--878.

\bibitem[\protect\citeauthoryear{Bulik-Sullivan, Loh, Finucane, Ripke, Yang,
  Patterson, Daly, Price, Neale, of~the Psychiatric Genomics~Consortium, {\it
  et~al.}}{\textsc{Bulik-Sullivan} {\em et~al.\@}}{2015}]{bulik2015ld}
\textsc{Bulik-Sullivan, B.~K.}, \textsc{P.-R. Loh}, \textsc{H.~K. Finucane},
  \textsc{S.~Ripke}, \textsc{J.~Yang}, \textsc{N.~Patterson}, \textsc{M.~J.
  Daly}, \textsc{A.~L. Price}, \textsc{B.~M. Neale}, \textsc{S.~W.~G. of~the
  Psychiatric Genomics~Consortium}, and \textsc{others}, 2015\ \ LD Score
  regression distinguishes confounding from polygenicity in genome-wide
  association studies.
\newblock Nature genetics~{\em 47\/}(3)\textbf{:}\ 291.

\bibitem[\protect\citeauthoryear{Bycroft, Freeman, Petkova, Band, Elliott,
  Sharp, Motyer, Vukcevic, Delaneau, O’Connell, {\it
  et~al.}}{\textsc{Bycroft} {\em et~al.\@}}{2018}]{bycroft2018uk}
\textsc{Bycroft, C.}, \textsc{C.~Freeman}, \textsc{D.~Petkova},
  \textsc{G.~Band}, \textsc{L.~T. Elliott}, \textsc{K.~Sharp},
  \textsc{A.~Motyer}, \textsc{D.~Vukcevic}, \textsc{O.~Delaneau},
  \textsc{J.~O’Connell}, and \textsc{others}, 2018\ \ The UK Biobank resource
  with deep phenotyping and genomic data.
\newblock Nature~{\em 562\/}(7726)\textbf{:}\ 203.

\bibitem[\protect\citeauthoryear{Carlson, Matise, North, Haiman, Fesinmeyer,
  Buyske, Schumacher, Peters, Franceschini, Ritchie, {\it
  et~al.}}{\textsc{Carlson} {\em et~al.\@}}{2013}]{carlson2013generalization}
\textsc{Carlson, C.~S.}, \textsc{T.~C. Matise}, \textsc{K.~E. North},
  \textsc{C.~A. Haiman}, \textsc{M.~D. Fesinmeyer}, \textsc{S.~Buyske},
  \textsc{F.~R. Schumacher}, \textsc{U.~Peters}, \textsc{N.~Franceschini},
  \textsc{M.~D. Ritchie}, and \textsc{others}, 2013\ \ Generalization and
  dilution of association results from European GWAS in populations of
  non-European ancestry: the PAGE study.
\newblock PLoS biology~{\em 11\/}(9)\textbf{:}\ e1001661.

\bibitem[\protect\citeauthoryear{Cole, Florez, and Hirschhorn}{\textsc{Cole}
  {\em et~al.\@}}{2019}]{cole2019comprehensive}
\textsc{Cole, J.~B.}, \textsc{J.~C. Florez}, and \textsc{J.~N. Hirschhorn},
  2019\ \ Comprehensive genomic analysis of dietary habits in UK Biobank
  identifies hundreds of genetic loci and establishes causal relationships
  between educational attainment and healthy eating.
\newblock BioRxiv\textbf{:}\ 662239.

\bibitem[\protect\citeauthoryear{Cox, Ruff, Maier, and Mathieson}{\textsc{Cox}
  {\em et~al.\@}}{2019}]{cox2019genetic}
\textsc{Cox, S.~L.}, \textsc{C.~B. Ruff}, \textsc{R.~M. Maier}, and
  \textsc{I.~Mathieson}, 2019\ \ Genetic contributions to variation in human
  stature in prehistoric Europe.
\newblock bioRxiv\textbf{:}\ 690545.

\bibitem[\protect\citeauthoryear{Crawford, Kelly, Hansen, Beltrame, Fan,
  Bowman, Jewett, Ranciaro, Thompson, Lo, {\it et~al.}}{\textsc{Crawford} {\em
  et~al.\@}}{2017}]{crawford2017loci}
\textsc{Crawford, N.~G.}, \textsc{D.~E. Kelly}, \textsc{M.~E. Hansen},
  \textsc{M.~H. Beltrame}, \textsc{S.~Fan}, \textsc{S.~L. Bowman},
  \textsc{E.~Jewett}, \textsc{A.~Ranciaro}, \textsc{S.~Thompson},
  \textsc{Y.~Lo}, and \textsc{others}, 2017\ \ Loci associated with skin
  pigmentation identified in African populations.
\newblock Science~{\em 358\/}(6365)\textbf{:}\ eaan8433.

\bibitem[\protect\citeauthoryear{Durvasula and Lohmueller}{\textsc{Durvasula}
  and \textsc{Lohmueller}}{2019}]{durvasula2019negative}
\textsc{Durvasula, A.} and \textsc{K.~E. Lohmueller}, 2019\ \ Negative
  selection on complex traits limits genetic risk prediction accuracy between
  populations.
\newblock bioRxiv\textbf{:}\ 721936.

\bibitem[\protect\citeauthoryear{Edwards, Bigham, Tan, Li, Gozdzik, Ross, Jin,
  and Parra}{\textsc{Edwards} {\em et~al.\@}}{2010}]{Edwards:2010kda}
\textsc{Edwards, M.}, \textsc{A.~Bigham}, \textsc{J.~Tan}, \textsc{S.~Li},
  \textsc{A.~Gozdzik}, \textsc{K.~Ross}, \textsc{L.~Jin}, and \textsc{E.~J.
  Parra}, 2010, April){Association of the OCA2 Polymorphism His615Arg with
  Melanin Content in East Asian Populations: Further Evidence of Convergent
  Evolution of Skin Pigmentation}.
\newblock PLoS Genetics~{\em 6\/}(3)\textbf{:}\ e1000867.

\bibitem[\protect\citeauthoryear{Feldman and Lewontin}{\textsc{Feldman} and
  \textsc{Lewontin}}{1975}]{feldman1975heritability}
\textsc{Feldman, M.~W.} and \textsc{R.~C. Lewontin}, 1975\ \ The heritability
  hang-up.
\newblock Science~{\em 190\/}(4220)\textbf{:}\ 1163--1168.

\bibitem[\protect\citeauthoryear{Fisher}{\textsc{Fisher}}{1935}]{fisher1937design}
\textsc{Fisher, R.~A.}, 1935\ \ {\em The design of experiments}.
\newblock Oliver And Boyd; Edinburgh; London.

\bibitem[\protect\citeauthoryear{Gianola}{\textsc{Gianola}}{1982}]{gianola1982assortative}
\textsc{Gianola, D.}, 1982\ \ Assortative mating and the genetic correlation.
\newblock Theoretical and Applied Genetics~{\em 62\/}(3)\textbf{:}\ 225--231.

\bibitem[\protect\citeauthoryear{Guo, Wu, Zhu, Zheng, Trzaskowski, Zeng,
  Robinson, Visscher, and Yang}{\textsc{Guo} {\em et~al.\@}}{2018}]{guo2018}
\textsc{Guo, J.}, \textsc{Y.~Wu}, \textsc{Z.~Zhu}, \textsc{Z.~Zheng},
  \textsc{M.~Trzaskowski}, \textsc{J.~Zeng}, \textsc{M.~R. Robinson},
  \textsc{P.~M. Visscher}, and \textsc{J.~Yang}, 2018\ \ Global genetic
  differentiation of complex traits shaped by natural selection in humans.
\newblock Nature communications~{\em 9\/}(1)\textbf{:}\ 1865.

\bibitem[\protect\citeauthoryear{Haworth, Mitchell, Corbin, Wade, Dudding,
  Budu-Aggrey, Carslake, Hemani, Paternoster, Smith, {\it
  et~al.}}{\textsc{Haworth} {\em et~al.\@}}{2019}]{haworth2019apparent}
\textsc{Haworth, S.}, \textsc{R.~Mitchell}, \textsc{L.~Corbin}, \textsc{K.~H.
  Wade}, \textsc{T.~Dudding}, \textsc{A.~Budu-Aggrey}, \textsc{D.~Carslake},
  \textsc{G.~Hemani}, \textsc{L.~Paternoster}, \textsc{G.~D. Smith}, and
  \textsc{others}, 2019\ \ Apparent latent structure within the UK Biobank
  sample has implications for epidemiological analysis.
\newblock Nature communications~{\em 10\/}(1)\textbf{:}\ 333.

\bibitem[\protect\citeauthoryear{Kerminen, Martin, Koskela, Ruotsalainen,
  Havulinna, Surakka, Palotie, Perola, Salomaa, Daly, {\it
  et~al.}}{\textsc{Kerminen} {\em et~al.\@}}{2018}]{kerminen2018geographic}
\textsc{Kerminen, S.}, \textsc{A.~R. Martin}, \textsc{J.~Koskela},
  \textsc{S.~E. Ruotsalainen}, \textsc{A.~S. Havulinna}, \textsc{I.~Surakka},
  \textsc{A.~Palotie}, \textsc{M.~Perola}, \textsc{V.~Salomaa}, \textsc{M.~J.
  Daly}, and \textsc{others}, 2018\ \ Geographic variation and bias in
  polygenic scores of complex diseases and traits in Finland.
\newblock bioRxiv\textbf{:}\ 485441.

\bibitem[\protect\citeauthoryear{Khera, Chaffin, Aragam, Haas, Roselli, Choi,
  Natarajan, Lander, Lubitz, Ellinor, {\it et~al.}}{\textsc{Khera} {\em
  et~al.\@}}{2018}]{khera2018genome}
\textsc{Khera, A.~V.}, \textsc{M.~Chaffin}, \textsc{K.~G. Aragam},
  \textsc{M.~E. Haas}, \textsc{C.~Roselli}, \textsc{S.~H. Choi},
  \textsc{P.~Natarajan}, \textsc{E.~S. Lander}, \textsc{S.~A. Lubitz},
  \textsc{P.~T. Ellinor}, and \textsc{others}, 2018\ \ Genome-wide polygenic
  scores for common diseases identify individuals with risk equivalent to
  monogenic mutations.
\newblock Nature genetics~{\em 50\/}(9)\textbf{:}\ 1219.

\bibitem[\protect\citeauthoryear{Kong, Thorleifsson, Frigge, Vilhjalmsson,
  Young, Thorgeirsson, Benonisdottir, Oddsson, Halldorsson, Masson, {\it
  et~al.}}{\textsc{Kong} {\em et~al.\@}}{2018}]{kong2018nature}
\textsc{Kong, A.}, \textsc{G.~Thorleifsson}, \textsc{M.~L. Frigge},
  \textsc{B.~J. Vilhjalmsson}, \textsc{A.~I. Young}, \textsc{T.~E.
  Thorgeirsson}, \textsc{S.~Benonisdottir}, \textsc{A.~Oddsson}, \textsc{B.~V.
  Halldorsson}, \textsc{G.~Masson}, and \textsc{others}, 2018\ \ The nature of
  nurture: Effects of parental genotypes.
\newblock Science~{\em 359\/}(6374)\textbf{:}\ 424--428.

\bibitem[\protect\citeauthoryear{Lander and Schork}{\textsc{Lander} and
  \textsc{Schork}}{1994}]{lander1994genetic}
\textsc{Lander, E.~S.} and \textsc{N.~J. Schork}, 1994\ \ Genetic dissection of
  complex traits.
\newblock Science~{\em 265\/}(5181)\textbf{:}\ 2037--2048.

\bibitem[\protect\citeauthoryear{Lawson, Davies, Haworth, Ashraf, Howe,
  Crawford, Hemani, Smith, and Timpson}{\textsc{Lawson} {\em
  et~al.\@}}{2019}]{lawson2019population}
\textsc{Lawson, D.~J.}, \textsc{N.~M. Davies}, \textsc{S.~Haworth},
  \textsc{B.~Ashraf}, \textsc{L.~Howe}, \textsc{A.~Crawford},
  \textsc{G.~Hemani}, \textsc{G.~D. Smith}, and \textsc{N.~J. Timpson}, 2019\ \
  Is population structure in the genetic biobank era irrelevant, a challenge,
  or an opportunity?
\newblock Human genetics\textbf{:}\ 1--19.

\bibitem[\protect\citeauthoryear{Marigorta and Navarro}{\textsc{Marigorta} and
  \textsc{Navarro}}{2013}]{marigorta2013high}
\textsc{Marigorta, U.~M.} and \textsc{A.~Navarro}, 2013\ \ High trans-ethnic
  replicability of GWAS results implies common causal variants.
\newblock PLoS genetics~{\em 9\/}(6)\textbf{:}\ e1003566.

\bibitem[\protect\citeauthoryear{Martin, Kanai, Kamatani, Okada, Neale, and
  Daly}{\textsc{Martin} {\em et~al.\@}}{2019}]{martin2019clinical}
\textsc{Martin, A.~R.}, \textsc{M.~Kanai}, \textsc{Y.~Kamatani},
  \textsc{Y.~Okada}, \textsc{B.~M. Neale}, and \textsc{M.~J. Daly}, 2019\ \
  Clinical use of current polygenic risk scores may exacerbate health
  disparities.
\newblock Nature genetics~{\em 51\/}(4)\textbf{:}\ 584.

\bibitem[\protect\citeauthoryear{Martin, Lin, Granka, Myrick, Liu, Sockell,
  Atkinson, Werely, M{\"o}ller, Sandhu, {\it et~al.}}{\textsc{Martin} {\em
  et~al.\@}}{2017}]{martin2017unexpectedly}
\textsc{Martin, A.~R.}, \textsc{M.~Lin}, \textsc{J.~M. Granka}, \textsc{J.~W.
  Myrick}, \textsc{X.~Liu}, \textsc{A.~Sockell}, \textsc{E.~G. Atkinson},
  \textsc{C.~J. Werely}, \textsc{M.~M{\"o}ller}, \textsc{M.~S. Sandhu}, and
  \textsc{others}, 2017\ \ An unexpectedly complex architecture for skin
  pigmentation in Africans.
\newblock Cell~{\em 171\/}(6)\textbf{:}\ 1340--1353.

\bibitem[\protect\citeauthoryear{Mathieson, Lazaridis, Rohland, Mallick,
  Patterson, Roodenberg, Harney, Stewardson, Fernandes, Novak, Sirak, Gamba,
  Jones, Llamas, Dryomov, Pickrell, Arsuaga, de~Castro, Carbonell, Gerritsen,
  Khokhlov, Kuznetsov, Lozano, Meller, Mochalov, Moiseyev, Guerra, Roodenberg,
  Verg{\`e}s, Krause, Cooper, Alt, Brown, Anthony, Lalueza-Fox, Haak, Pinhasi,
  and Reich}{\textsc{Mathieson} {\em et~al.\@}}{2015}]{Mathieson:2015cy}
\textsc{Mathieson, I.}, \textsc{I.~Lazaridis}, \textsc{N.~Rohland},
  \textsc{S.~Mallick}, \textsc{N.~Patterson}, \textsc{S.~A. Roodenberg},
  \textsc{E.~Harney}, \textsc{K.~Stewardson}, \textsc{D.~Fernandes},
  \textsc{M.~Novak}, \textsc{K.~Sirak}, \textsc{C.~Gamba}, \textsc{E.~R.
  Jones}, \textsc{B.~Llamas}, \textsc{S.~Dryomov}, \textsc{J.~Pickrell},
  \textsc{J.~L. Arsuaga}, \textsc{J.~M.~B. de~Castro}, \textsc{E.~Carbonell},
  \textsc{F.~Gerritsen}, \textsc{A.~Khokhlov}, \textsc{P.~Kuznetsov},
  \textsc{M.~Lozano}, \textsc{H.~Meller}, \textsc{O.~Mochalov},
  \textsc{V.~Moiseyev}, \textsc{M.~A.~R. Guerra}, \textsc{J.~Roodenberg},
  \textsc{J.~M. Verg{\`e}s}, \textsc{J.~Krause}, \textsc{A.~Cooper},
  \textsc{K.~W. Alt}, \textsc{D.~Brown}, \textsc{D.~Anthony},
  \textsc{C.~Lalueza-Fox}, \textsc{W.~Haak}, \textsc{R.~Pinhasi}, and
  \textsc{D.~Reich}, 2015, December){Genome-wide patterns of selection in 230
  ancient Eurasians}.
\newblock Nature~{\em 528\/}(7583)\textbf{:}\ 499--503.

\bibitem[\protect\citeauthoryear{Mostafavi, Harpak, Conley, Pritchard, and
  Przeworski}{\textsc{Mostafavi} {\em et~al.\@}}{2019}]{mostafavi2019variable}
\textsc{Mostafavi, H.}, \textsc{A.~Harpak}, \textsc{D.~Conley}, \textsc{J.~K.
  Pritchard}, and \textsc{M.~Przeworski}, 2019\ \ Variable prediction accuracy
  of polygenic scores within an ancestry group.
\newblock BioRxiv\textbf{:}\ 629949.

\bibitem[\protect\citeauthoryear{North}{\textsc{North}}{2015}]{CuppaTheGrocer}
\textsc{North, A.}, 2015\ \ What your cuppa says about you.
\newblock {The Grocer}.

\bibitem[\protect\citeauthoryear{Norton, Kittles, Parra, McKeigue, Mao, Cheng,
  Canfield, Bradley, McEvoy, and Shriver}{\textsc{Norton} {\em
  et~al.\@}}{2006}]{Norton:2006kpa}
\textsc{Norton, H.~L.}, \textsc{R.~A. Kittles}, \textsc{E.~Parra},
  \textsc{P.~McKeigue}, \textsc{X.~Mao}, \textsc{K.~Cheng}, \textsc{V.~A.
  Canfield}, \textsc{D.~G. Bradley}, \textsc{B.~McEvoy}, and \textsc{M.~D.
  Shriver}, 2006, December){Genetic Evidence for the Convergent Evolution of
  Light Skin in Europeans and East Asians}.
\newblock Molecular Biology and Evolution~{\em 24\/}(3)\textbf{:}\ 710--722.

\bibitem[\protect\citeauthoryear{Olalde, Allentoft, S{\'a}nchez-Quinto,
  Santpere, Chiang, DeGiorgio, Prado-Martinez, Rodr{\'\i}guez, Rasmussen,
  Quilez, {\it et~al.}}{\textsc{Olalde} {\em
  et~al.\@}}{2014}]{olalde2014derived}
\textsc{Olalde, I.}, \textsc{M.~E. Allentoft}, \textsc{F.~S{\'a}nchez-Quinto},
  \textsc{G.~Santpere}, \textsc{C.~W. Chiang}, \textsc{M.~DeGiorgio},
  \textsc{J.~Prado-Martinez}, \textsc{J.~A. Rodr{\'\i}guez},
  \textsc{S.~Rasmussen}, \textsc{J.~Quilez}, and \textsc{others}, 2014\ \
  Derived immune and ancestral pigmentation alleles in a 7,000-year-old
  Mesolithic European.
\newblock Nature~{\em 507\/}(7491)\textbf{:}\ 225.

\bibitem[\protect\citeauthoryear{Orwell}{\textsc{Orwell}}{1946}]{Orwell:1946}
\textsc{Orwell, G.}, 1946\ \ A Nice Cup of Tea.
\newblock London Evening Standard.

\bibitem[\protect\citeauthoryear{Penrose}{\textsc{Penrose}}{1955}]{penrose1955heredity}
\textsc{Penrose, L.~S.}, 1955\ \ {\em Heredity and environment in human
  affairs}.
\newblock National Children's Home.

\bibitem[\protect\citeauthoryear{Price, Zaitlen, Reich, and
  Patterson}{\textsc{Price} {\em et~al.\@}}{2010}]{price2010new}
\textsc{Price, A.~L.}, \textsc{N.~A. Zaitlen}, \textsc{D.~Reich}, and
  \textsc{N.~Patterson}, 2010\ \ New approaches to population stratification in
  genome-wide association studies.
\newblock Nature Reviews Genetics~{\em 11\/}(7)\textbf{:}\ 459.

\bibitem[\protect\citeauthoryear{Racimo, Berg, and Pickrell}{\textsc{Racimo}
  {\em et~al.\@}}{2018}]{Racimo18}
\textsc{Racimo, F.}, \textsc{J.~J. Berg}, and \textsc{J.~K. Pickrell}, 2018\ \
  Detecting polygenic adaptation in admixture graphs.
\newblock Genetics~{\em 208\/}(4)\textbf{:}\ 1565--1584.

\bibitem[\protect\citeauthoryear{Robinson, Hemani, Medina-Gomez, Mezzavilla,
  Esko, Shakhbazov, Powell, Vinkhuyzen, Berndt, Gustafsson, Justice, Kahali,
  Locke, Pers, Vedantam, Wood, van Rheenen, Andreassen, Gasparini, Metspalu,
  van~den Berg, Veldink, Rivadeneira, Werge, Abecasis, Boomsma, Chasman,
  de~Geus, Frayling, Hirschhorn, Hottenga, Ingelsson, Loos, Magnusson, Martin,
  Montgomery, North, Pedersen, Spector, Speliotes, Goddard, Yang, and
  Visscher}{\textsc{Robinson} {\em et~al.\@}}{2015}]{Robinson:2015ki}
\textsc{Robinson, M.~R.}, \textsc{G.~Hemani}, \textsc{C.~Medina-Gomez},
  \textsc{M.~Mezzavilla}, \textsc{T.~Esko}, \textsc{K.~Shakhbazov},
  \textsc{J.~E. Powell}, \textsc{A.~Vinkhuyzen}, \textsc{S.~I. Berndt},
  \textsc{S.~Gustafsson}, \textsc{A.~E. Justice}, \textsc{B.~Kahali},
  \textsc{A.~E. Locke}, \textsc{T.~H. Pers}, \textsc{S.~Vedantam},
  \textsc{A.~R. Wood}, \textsc{W.~van Rheenen}, \textsc{O.~A. Andreassen},
  \textsc{P.~Gasparini}, \textsc{A.~Metspalu}, \textsc{L.~H. van~den Berg},
  \textsc{J.~H. Veldink}, \textsc{F.~Rivadeneira}, \textsc{T.~M. Werge},
  \textsc{G.~R. Abecasis}, \textsc{D.~I. Boomsma}, \textsc{D.~I. Chasman},
  \textsc{E.~J.~C. de~Geus}, \textsc{T.~M. Frayling}, \textsc{J.~N.
  Hirschhorn}, \textsc{J.~J. Hottenga}, \textsc{E.~Ingelsson}, \textsc{R.~J.~F.
  Loos}, \textsc{P.~K.~E. Magnusson}, \textsc{N.~G. Martin}, \textsc{G.~W.
  Montgomery}, \textsc{K.~E. North}, \textsc{N.~L. Pedersen}, \textsc{T.~D.
  Spector}, \textsc{E.~K. Speliotes}, \textsc{M.~E. Goddard}, \textsc{J.~Yang},
  and \textsc{P.~M. Visscher}, 2015, September){Population genetic
  differentiation of height and body mass index across Europe}.
\newblock Nature Publishing Group~{\em 47\/}(11)\textbf{:}\ 1357--1362.

\bibitem[\protect\citeauthoryear{Rosenberg, Edge, Pritchard, and
  Feldman}{\textsc{Rosenberg} {\em et~al.\@}}{2018}]{rosenberg2018interpreting}
\textsc{Rosenberg, N.~A.}, \textsc{M.~D. Edge}, \textsc{J.~K. Pritchard}, and
  \textsc{M.~W. Feldman}, 2018\ \ Interpreting polygenic scores, polygenic
  adaptation, and human phenotypic differences.
\newblock Evolution, medicine, and public health~{\em 2019\/}(1)\textbf{:}\
  26--34.

\bibitem[\protect\citeauthoryear{Salsburg}{\textsc{Salsburg}}{2001}]{salsburg2001lady}
\textsc{Salsburg, D.}, 2001\ \ {\em The lady tasting tea: How statistics
  revolutionized science in the twentieth century}.
\newblock Macmillan.

\bibitem[\protect\citeauthoryear{Sohail, Maier, Ganna, Bloemendal, Martin,
  Turchin, Chiang, Hirschhorn, Daly, Patterson, {\it et~al.}}{\textsc{Sohail}
  {\em et~al.\@}}{2019}]{sohail2019polygenic}
\textsc{Sohail, M.}, \textsc{R.~M. Maier}, \textsc{A.~Ganna},
  \textsc{A.~Bloemendal}, \textsc{A.~R. Martin}, \textsc{M.~C. Turchin},
  \textsc{C.~W. Chiang}, \textsc{J.~Hirschhorn}, \textsc{M.~J. Daly},
  \textsc{N.~Patterson}, and \textsc{others}, 2019\ \ Polygenic adaptation on
  height is overestimated due to uncorrected stratification in genome-wide
  association studies.
\newblock eLife~{\bf 8}\textbf{:}\ e39702.

\bibitem[\protect\citeauthoryear{Tabery}{\textsc{Tabery}}{2008}]{tabery2008ra}
\textsc{Tabery, J.}, 2008\ \ RA Fisher, Lancelot Hogben, and the origin (s) of
  genotype--environment interaction.
\newblock Journal of the History of Biology~{\em 41\/}(4)\textbf{:}\ 717--761.

\bibitem[\protect\citeauthoryear{Taylor, Smith, and Munaf{\`o}}{\textsc{Taylor}
  {\em et~al.\@}}{2018}]{taylor2018associations}
\textsc{Taylor, A.~E.}, \textsc{G.~D. Smith}, and \textsc{M.~R. Munaf{\`o}},
  2018\ \ Associations of coffee genetic risk scores with consumption of
  coffee, tea and other beverages in the UK Biobank.
\newblock Addiction~{\em 113\/}(1)\textbf{:}\ 148--157.

\bibitem[\protect\citeauthoryear{Turchin, Chiang, Palmer, Sankararaman, Reich,
  and Hirschhorn}{\textsc{Turchin} {\em et~al.\@}}{2012}]{Turchin:2012dy}
\textsc{Turchin, M.~C.}, \textsc{C.~W. Chiang}, \textsc{C.~D. Palmer},
  \textsc{S.~Sankararaman}, \textsc{D.~Reich}, and \textsc{J.~N. Hirschhorn},
  2012, August){Evidence of widespread selection on standing variation in
  Europe at height-associated SNPs}.
\newblock Nature Publishing Group~{\em 44\/}(9)\textbf{:}\ 1015--1019.

\bibitem[\protect\citeauthoryear{Vilhj{\'a}lmsson and
  Nordborg}{\textsc{Vilhj{\'a}lmsson} and
  \textsc{Nordborg}}{2012}]{vilhjalmsson2012nature}
\textsc{Vilhj{\'a}lmsson, B.~J.} and \textsc{M.~Nordborg}, 2012\ \ The nature
  of confounding in genome-wide association studies.
\newblock Nature Reviews Genetics~{\em 14\/}(1)\textbf{:}\ 1.

\bibitem[\protect\citeauthoryear{Visscher, Wray, Zhang, Sklar, McCarthy, Brown,
  and Yang}{\textsc{Visscher} {\em et~al.\@}}{2017}]{visscher201710}
\textsc{Visscher, P.~M.}, \textsc{N.~R. Wray}, \textsc{Q.~Zhang},
  \textsc{P.~Sklar}, \textsc{M.~I. McCarthy}, \textsc{M.~A. Brown}, and
  \textsc{J.~Yang}, 2017\ \ 10 years of GWAS discovery: biology, function, and
  translation.
\newblock The American Journal of Human Genetics~{\em 101\/}(1)\textbf{:}\
  5--22.

\bibitem[\protect\citeauthoryear{Williams and Jackson}{\textsc{Williams} and
  \textsc{Jackson}}{2005}]{williams2005social}
\textsc{Williams, D.~R.} and \textsc{P.~B. Jackson}, 2005\ \ Social sources of
  racial disparities in health.
\newblock Health affairs~{\em 24\/}(2)\textbf{:}\ 325--334.

\bibitem[\protect\citeauthoryear{Williams and Mohammed}{\textsc{Williams} and
  \textsc{Mohammed}}{2009}]{williams2009discrimination}
\textsc{Williams, D.~R.} and \textsc{S.~A. Mohammed}, 2009\ \ Discrimination
  and racial disparities in health: evidence and needed research.
\newblock Journal of behavioral medicine~{\em 32\/}(1)\textbf{:}\ 20--47.

\bibitem[\protect\citeauthoryear{Zoledziewska, Sidore, Chiang, Sanna, Mulas,
  Steri, Busonero, Marcus, Marongiu, Maschio, Del~Vecchyo, Floris, Meloni,
  Delitala, Concas, Murgia, Biino, Vaccargiu, Nagaraja, Lohmueller, Timpson,
  Soranzo, Tachmazidou, Dedoussis, Zeggini, Uzzau, Jones, Lyons, Angius,
  Abecasis, Novembre, Schlessinger, and Cucca}{\textsc{Zoledziewska} {\em
  et~al.\@}}{2015}]{Zoledziewska:2015do}
\textsc{Zoledziewska, M.}, \textsc{C.~Sidore}, \textsc{C.~W.~K. Chiang},
  \textsc{S.~Sanna}, \textsc{A.~Mulas}, \textsc{M.~Steri},
  \textsc{F.~Busonero}, \textsc{J.~H. Marcus}, \textsc{M.~Marongiu},
  \textsc{A.~Maschio}, \textsc{D.~O. Del~Vecchyo}, \textsc{M.~Floris},
  \textsc{A.~Meloni}, \textsc{A.~Delitala}, \textsc{M.~P. Concas},
  \textsc{F.~Murgia}, \textsc{G.~Biino}, \textsc{S.~Vaccargiu},
  \textsc{R.~Nagaraja}, \textsc{K.~E. Lohmueller}, \textsc{N.~J. Timpson},
  \textsc{N.~Soranzo}, \textsc{I.~Tachmazidou}, \textsc{G.~Dedoussis},
  \textsc{E.~Zeggini}, \textsc{S.~Uzzau}, \textsc{C.~Jones}, \textsc{R.~Lyons},
  \textsc{A.~Angius}, \textsc{G.~R. Abecasis}, \textsc{J.~Novembre},
  \textsc{D.~Schlessinger}, and \textsc{F.~Cucca}, 2015,
  September){Height-reducing variants and selection for short stature in
  Sardinia}.
\newblock Nature Publishing Group~{\em 47\/}(11)\textbf{:}\ 1352--1356.

\end{thebibliography}

%\bibliography{pnas-sample}

\end{document}